\documentclass[sigconf]{acmart}
\usepackage{caption}
\usepackage{subcaption}
\usepackage{dblfloatfix}
\usepackage[utf8]{inputenc}

\AtBeginDocument{%
  \providecommand\BibTeX{{%
    \normalfont B\kern-0.5em{\scshape i\kern-0.25em b}\kern-0.8em\TeX}}}

\setcopyright{acmcopyright}
\copyrightyear{2023}
\acmYear{2023}
\acmDOI{XXXXXXX.XXXXXXX}

\acmConference[SIGSPATIAL '23]{the 31st ACM SIGSPATIAL
International Conference on Advances
in Geographic Information Systems}{November 13--16,
  2023}{Hamburg, Germany}
\acmPrice{}
\acmISBN{978-1-4503-XXXX-X/18/06}

\usepackage{psboxit}

\begin{document}
\title{On the Use of Abundant Road Speed Data for Travel Demand Calibration of Urban Traffic Simulators}
\author{Suyash Vishnoi}
\authornote{Work done while at Google Research}
\email{scvishnoi@google.com}
\orcid{0000-0002-6592-818X}
\affiliation{%
  \institution{Google Research/ UT Austin}
  \streetaddress{}
  \city{Mountain View}
  \state{CA}
  \country{USA/ Austin, TX, USA}
  \postcode{}
}

\author{Akhil Shetty}
\authornotemark[1]
\email{shetty.akhil@berkeley.edu}
\affiliation{%
  \institution{UC Berkeley}
  \streetaddress{}
  \city{Berkeley}
  \state{CA}
  \country{USA}
  \postcode{}
}

\author{Iveel Tsogsuren}
\email{iveel@google.com}
\affiliation{%
  \institution{Google Research}
  \streetaddress{}
  \city{Mountain View}
  \state{CA}
  \country{USA}
  \postcode{}
}

\author{Neha Arora}
\email{nehaarora@google.com}
\affiliation{%
  \institution{Google Research}
  \streetaddress{}
  \city{Mountain View}
  \state{CA}
  \country{USA}
  \postcode{}
}

\author{Carolina Osorio}
\authornotemark[1]
\email{osorioc@google.com}
\affiliation{%
  \institution{Google Research/ HEC Montreal}
  \streetaddress{}
  \city{Mountain View}
  \state{CA}
  \country{USA/Montreal, QC, Canada}
  \postcode{}
}

\renewcommand{\shortauthors}{Vishnoi, Shetty, Tsogsuren, Arora and Osorio}


\begin{abstract}
  This work develops a compute-efficient algorithm to tackle a fundamental problem in transportation: that of urban travel demand estimation. It focuses on the calibration of origin-destination travel demand input parameters for high-resolution traffic simulation models. It considers the use of abundant traffic road speed data. The travel demand calibration problem is formulated as a continuous, high-dimensional, simulation-based optimization (SO) problem with bound constraints. There is a lack of compute efficient algorithms to tackle this problem. We propose the use of an SO algorithm that relies on an efficient, analytical, differentiable, physics-based traffic model, known as a metamodel or surrogate model. We formulate a metamodel that enables the use of road speed data. Tests are performed on a Salt Lake City network. We study how the amount of data, as well as the congestion levels, impact both in-sample and out-of-sample performance.
  The proposed method outperforms the benchmark  for both in-sample and out-of-sample performance by $84.4\%$ and $72.2\%$ in terms of speeds and counts, respectively. Most importantly, the proposed method yields the highest compute efficiency, identifying solutions with good performance within few simulation function evaluations (i.e., with small samples). 
\end{abstract}



\received{}
\received[revised]{}
\received[accepted]{}

\maketitle
\section{Introduction}
Traffic microsimulation software is a widely used tool to plan changes in the city by simulating the impact of those changes before implementing them in the real-world. To assess the impact of large-scale changes it is important that the  simulator be well calibrated to the traffic conditions of the actual network.
These settings include the Origin-Destination (OD) demands which determines the amount of traffic flows on the network. 
Most often, the ground truth (GT) data available for OD calibration is spatially sparse  vehicular count data on segments  \cite{osorio2019high}. Work that includes GT
speed data include \cite{cipriani2011gradient,huang2010accelerated}.
In this paper, we focus on the use of non-spatially sparse segment (i.e., a road/link) speed data for travel demand calibration of traffic microsimulation models. An efficient method to solve the proposed optimization problem is presented and compared with an existing algorithm for speeds-based calibration from the literature. 

Demand calibration problems of microsimulation models are simulation-based, stochastic, nonlinear and high dimensional problems. A sample efficient algorithm for OD demand calibration using vehicle count data on large-scale networks is presented in \cite{osorio2019high} which exploits a physics-based metamodel to perform efficient calibration. This approach uses a low-resolution analytical traffic model to convert the OD demands which are the features of the problem to segment counts and further uses a statistical model as simple as a linear regression model to learn the difference between the counts generated by the analytical traffic model and the simulation. This results in a loss function that is differentiable making it compatible with efficient  gradient-based solvers. 
%
The present work extends the approach from \cite{osorio2019high} to speeds-based OD demand calibration for urban road networks, while making the following contributions.

We propose a novel metamodel formulation that is scalable and computationally efficient. It is based on a novel relationship between segment demands and segment speeds \cite{lia2022fundamental}, known as the fundamental diagram (FD), in calibration which is appropriate for arterial/urban segments compared to existing studies \cite{osorio2019high}.
 The proposed approach is tested under different levels of sparsity of segments with GT speed data to assess the value of the amount of speed data in calibration. 

\section{Methodology}
\label{s:methodology}

An urban area is discretized into traffic analysis zones (TAZs), which can serve as origins or destinations for trips. Pairs of TAZs are referred to as origin-destination (OD) pairs. In this work, we focus on the offline calibration of demands for a subset of all OD pairs such that the simulated traffic metrics are close to the ground truth (GT) metrics. The problem considered in this work can be defined as follows:

\textbf{Problem 1.} \textit{Given the GT speeds for a set of segments on the network, what are the demands that can be loaded onto the OD pairs in a microsimulation model?}

The above problem can be written as an SO problem with the OD pair demands as features and the error between the simulated speeds and the GT speeds as the loss function. A detailed description of the optimization problem is given below.

To formulate the calibration problem, we introduce the following notation:

\begin{tabular}{l}
$x$:                 feature vector (i.e., vector of hourly OD demands),\\
$\mathcal{I}$:      set of segments with GT speeds,\\
$f(x)$:              simulation-based loss function,\\
$u_1$:                 vector of endogenous simulation variables,\\
$u_2$:                 vector of exogenous simulation parameters,\\
$v_i^{\text{GT}}$:               GT measurement of average (space mean) speed of segment $i$,\\
$E[v_i(x,u_1;u_2)]$:       expected (space mean) speed of segment $i$,\\
$x_U$:               vector of upper bounds on the hourly OD demands,\\
$v_i^{\text{max}}$:           speed limit of segment $i$,\\
$w_i$:		loss function weight corresponding to segment $i$.\\
\end{tabular}

The optimization problem is formulated as follows:
\begin{align}
\min_{x} \quad & f(x)=\frac{1}{|\mathcal{I}|}\sum_{i\in \mathcal{I}}w_i(v_i^{\text{GT}} - E[v_i(x,u_1;u_2)])^2\label{e:so_optimization_problem_loss} \\
\textrm{s.t.} \quad & 0 \le x \le x_U\label{e:so_optimization_problem_constraints}
\end{align}
Here, $|\mathcal{I}|$ denotes the cardinality of the set $\mathcal{I}$. The loss function of the problem is the minimization of the weighted mean squared error between the GT measurements of speeds and the corresponding expected speeds $E[v_i(x,u_1;u_2)]$ from the simulator. The corresponding segment weight $w_i$ in this work is defined as:
\begin{equation*}
w_i  = \min\biggl\{\frac{v_i^{\text{GT}}}{v_i^{\text{max}}}, 1-\frac{v_i^{\text{GT}}}{v_i^{\text{max}}}\biggl\}.
\end{equation*}
This weight term represents the importance of a segment for calibration which in this case implies that we put a lower weight on segments for which the GT speeds are either very close to the speed limit or are very close to 0 m/s. This is because traffic in these regimes is known to have a many-to-one relationship between counts and speeds (and hence demands and speeds) \cite{lia2022fundamental} which makes them less valuable than other segments which have a unique relationship between demands and speeds and thus drive the solver to a unique solution. Besides the feature vector, the expected space-mean speed of segments depend upon the endogenous variables $u_1$ such as queue lengths and travel time, and exogenous parameters $u_2$ such as segment attributes, car-following parameters, network topology, and route-choice probabilities among other parameters. Unlike \cite{osorio2019high}, here we do not consider a regularization term in the formulation.
\subsection*{Physics-informed Metamodel Algorithm}
\label{s:metamodel_based_solution}
To solve the optimization problem presented in \eqref{e:so_optimization_problem_loss}-\eqref{e:so_optimization_problem_constraints}, we propose a physics-informed metamodel algorithm inspired by \cite{osorio2019high} and \cite{osorio2019efficient}. This approach approximates the simulation-based and non-differentiable optimization problem loss function \eqref{e:so_optimization_problem_loss} with an analytical and differentiable metamodel, or surrogate model. The chosen metamodel is a combination of an analytical component relating the features with the loss function through a known physics-based or traffic-based  relationship, and a general-purpose, traffic agnostic, functional component which is a statistical model (in this case a linear model) that aims to learn the difference between the analytical physics-based model and the simulator through the feature vector. In simpler terms, the functional component is trained on the difference between the loss function calculated using the analytical physics-based model and that calculated using the simulator. The resulting problem is differentiable and can be solved with the help of efficient gradient-based algorithms. The metamodel algorithm is an iterative approach: at every epoch of the SO algorithm, new simulation function evaluations become available, the metamodel parameters are then trained or updated, so as to bring the metamodel closer to the actual simulation-based loss function. The newly trained metamodel is then used to solve an analytical and differentiable optimization problem, the solution of which is evaluated via simulation, leading to new simulation function evaluations. 
\begin{table*}[!b]
    \centering
    \vspace*{3mm}
    \caption{Average nRMSE over 5 runs of speeds-based calibration using the metamodel-based approach and SPSA along with that obtained using the initial demands (marked with `-' in the algorithm column).}
    \begin{tabular}{p{0.1\textwidth}p{0.12\textwidth}ccccc}
    \toprule
        \#in-sample segments & \#out-of-sample segments & Algorithm & \multicolumn{2}{c}{In-sample nRMSE, evaluated over:} & \multicolumn{2}{c}{Out-of-sample nRMSE, evaluated over:}\\
         & & & speeds & counts & speeds & counts\\
         \midrule
        331 & 1,951 & metamodel-based & 0.346 & 0.603 & 0.212& 0.884  \\
        & & SPSA & 0.593  & 1.118  & 0.396 & 1.534 \\
        & & - & 0.718 & 1.231 & 0.438 & 1.72\\
        456 & 1,826 & metamodel-based & 0.171  & 0.372  & 0.048 & 0.488 \\
         & & SPSA &  0.542 & 1.028 & 0.380 & 1.356\\
         & & - &  0.628 & 1.304 & 0.438 & 1.735\\
        2,282 & 0 & metamodel-based & 0.049  & 0.387  & - & - \\
         & & SPSA & 0.315  & 1.394  & - & -\\
         & & - & 0.461 & 1.624 & - & -\\
         \bottomrule
    \end{tabular}
    \label{t:nrmse_insample_speeds}
\end{table*}
The optimization problem solved at every epoch is known as the metamodel optimization problem.  To formulate it, we define the following notation.
\begin{tabular}{l}
$k$:                   algorithm epoch,\\
$m_k$:               metamodel function at epoch $k$,\\
$\beta_k$:           parameter vector of metamodel $m_k$,\\
$\beta_{k,j}$:       scalar element $j$ of the parameter vector,\\
$f_A(x)$:           physics-based analytical approximation of the \\optimization problem loss function \eqref{e:so_optimization_problem_loss} provided by an \\analytical traffic model,\\
$\phi(x;\beta_k)$:   functional component of the metamodel $m_k$,\\
$x_z$:               expected hourly demand for OD pair $z$,\\
$\mathcal{Z}$:       set of indices of OD pairs $\mathcal{Z}=\{1,2,\dots,z_0\}$,\\

\textbf{Endogenous variables of the analytical traffic model}\\
$v_i^{\text{a}}$:   physics-based analytical speed on segment $i$,\\
$q_i$:              expected hourly demand for segment $i$,\\
$q$:                 vector of expected hourly demands,\\

\textbf{Exogenous parameters of the analytical traffic model}\\
$A$:               matrix to map OD demands to segment demands,\\
$\alpha_{i}^1, \alpha_{i}^2$:      FD parameters for segment $i$,\\
$q^{\text{max}}_i$:         maximum demand for segment $i$,\\
$v^{\text{min}}_i$:         minimum speed on segment $i$.\\
\end{tabular}

The metamodel optimization problem at any epoch $k$ is written as follows:
\begin{align}
\min_{x} \quad & m_k(x;\beta_k)=\beta_{k,0}f_A(x) + \phi(x;
\beta_k)\label{e:metamodel}\\
\textrm{s.t.} \quad & f_A(x) = \frac{1}{|\mathcal{I}|}\sum_{i\in\mathcal{I}}{w_i(v_i^{\text{GT}}-v_i^{\text{a}})^2}\label{e:analytical}\\
\quad & \phi(x;\beta_k)=\beta_{k,1}+\sum_{z\in\mathcal{Z}}{\beta_{k,z+1}x_z}\label{e:functional}\\
\quad & v_i^{\text{a}} = v^{\text{min}}_i + (v^{\text{max}}_i - v^{\text{min}}_i)\left(1 - \left(\frac{q_i}{q_i^{\text{max}}}\right)^{\alpha^1_i}\right)^{\alpha^2_i}\forall i \in \mathcal{I}\label{e:urbanfd}\\
\quad & q = Ax\label{e:assignment}\\
\quad & 0 \le x \le x_U
\end{align}

The physics-based analytical component \eqref{e:analytical} of the metamodel is defined as the weighted mean squared error between the ground truth (GT) speeds and the physics-based analytical speeds, corresponding to the first expression in the loss function \eqref{e:metamodel}. The physics-based analytical model is defined using a recently proposed FD for urban segments \eqref{e:urbanfd} in \cite{lia2022fundamental} , which relates the segment speeds to segment demands, and \eqref{e:assignment}, which relates the segment demands to OD demands. Therefore, the physics-based analytical component gives an approximate analytical relationship between the feature vector which consists of the OD demands and the loss function of the SO problem \eqref{e:so_optimization_problem_loss}. In this work, we assume that the route choices are exogenous and independent of the actual flow of traffic at any epoch. Therefore, $A$ which is also known as the assignment matrix is fixed. Out of the parameters of the FD model, $v^{\text{max}}_i$, $v^{\text{min}}_i$, and $q^{\text{max}}_i$ are considered fixed for each segment and set based on observed GT data. On the other hand $\alpha_{i}^1$ and $\alpha_{i}^2$ are either fitted based on simulated traffic data for the network or assigned according to observed values from the literature \cite{lia2022fundamental}. 
The functional component $\phi(x,\beta_k)$ is a linear model defined using the OD demands. 
More details on metamodel training can be found in \cite{osorio2019high}. 
The bound constraints are the same as in the original problem \eqref{e:so_optimization_problem_constraints}.


\section{Salt Lake City Case Study}
\label{s:numerical_study}

\label{s:network_description}

This study uses a network model of Salt Lake City, Utah, created with SUMO \cite{sumo}. The network has 15,254 segments, including 2,282 major arterial segments on which traffic is observed in this study. It also has 2,818 intersections, of which 94 are signalized. We consider 62 origin-destination (OD) pairs in the network. Each of the 62 ODs has three possible routes, resulting in a total of 186 routes across the network. We use a synthetic OD demand set as ground truth (GT) demands. Demands are generated for the 62 OD pairs similar to \cite{mladenov2022adversarial}. GT field data is obtained from 10 simulation runs using the GT OD demands. We assume that no historical data is available about the GT demands. Therefore, we initiate the algorithm with a uniformly sampled random set of OD demands. The primary evaluation metric used to judge the quality of the calibrated demands is the normalized-\textit{root mean squared error} (nRMSE) \cite{arora2021efficient}.

We benchmark our speeds-based OD demand calibration approach against SPSA algorithm~\cite{spall2005introduction} on 3 sets of GT segments grouped by congestion level in a synthetic GT simulation:
\begin{itemize}
    \item 331 segments with speed/speed-limit $\leq$ 0.8
    \item 456 segments with speed/speed-limit $\leq$ 0.9 
    \item 2,282 segments with speed/speed-limit $\leq$ 1.0
\end{itemize}
For each set of GT segments, we run each algorithm 5 times. An algorithm run is terminated once a total of 250 ODs have been simulated. Final nRMSE estimated are obtained by averaging over 5 simulation replications. 

\begin{figure}
\begin{subfigure}{0.45\columnwidth}
\includegraphics[width=\textwidth]{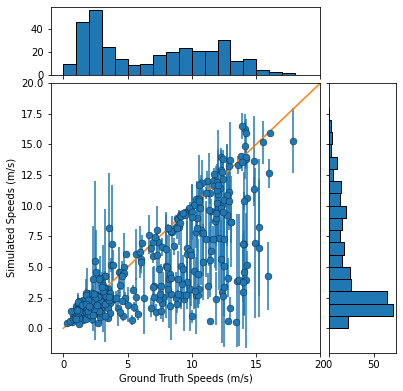}
\end{subfigure}
\hfill
\begin{subfigure}{0.45\columnwidth}
\includegraphics[width=\textwidth]{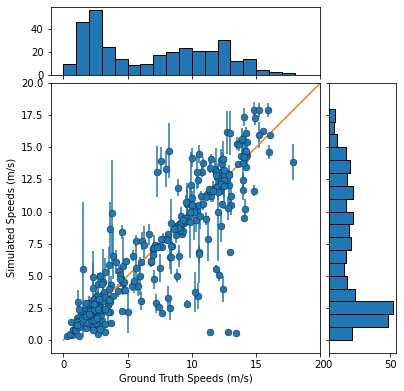}
\end{subfigure} 
\begin{subfigure}{0.45\columnwidth}
\includegraphics[width=\textwidth]{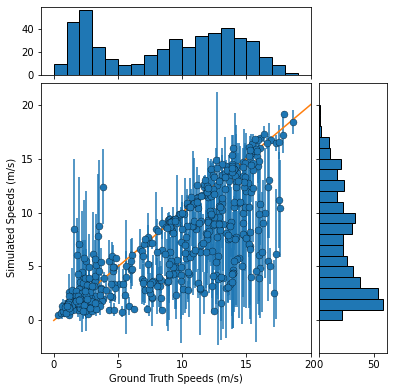} 
\end{subfigure}  
\hfill 
\begin{subfigure}{0.45\columnwidth} 
\includegraphics[width=\textwidth]{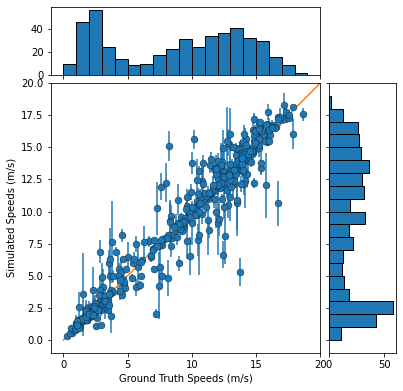} 
\end{subfigure}
\begin{subfigure}{0.45\columnwidth} 
\includegraphics[width=\textwidth]{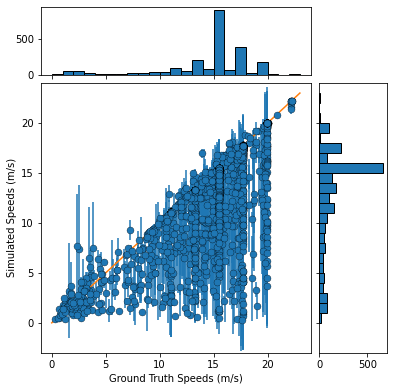} 
\end{subfigure}  
\hfill 
\begin{subfigure}{0.45\columnwidth} 
\includegraphics[width=\textwidth]{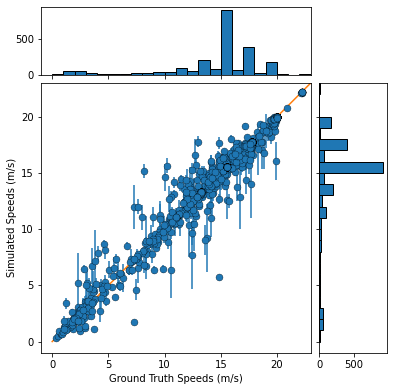} 
\end{subfigure}
\caption{Simulated vs. GT speeds for SPSA (left) and metamodel (right) calibrated demands. Top, middle, and bottom rows show results for calibration with 331, 456, and 2,282 segments, respectively. The $x=y$ line is shown for reference.}
\label{f:insample_nrmse_speeds}
\end{figure}

%
%
%
Plots of simulated speeds from calibrated demand against GT speeds for each segment set used for calibration are shown in Figure \ref{f:insample_nrmse_speeds}. Each point represents a segment, with the y-axis denoting average simulated speed over 5 calibration runs and the x-axis denoting GT speed. Error bars have a half-width of one standard deviation in simulated speeds over the 5 runs. Ideally, after calibration, simulated speeds should equal GT speeds. The top histogram shows the number of segments around the GT speeds on the x-axis, and the right histogram shows the number of segments with simulated speeds around the values on the y-axis. In, Figure \ref{f:insample_nrmse_speeds}, we compare the plots of columns 1 (SPSA plots) and 2 (metamodel plots) and it shows that the proposed approach outperforms SPSA in terms of  fit to GT data. This holds for all three sets of segments (i.e., all three rows of plots).

Table \ref{t:nrmse_insample_speeds} shows the in-sample and out-of-sample average nRMSE in speeds and counts for each segment set used for calibration, as well as the average nRMSE for initial demands. The in-sample nRMSE in speeds is the primary measure of convergence, as it is directly related to the metric minimized in the optimization problem. The average out-of-sample nRMSE in speeds and in-sample and out-of-sample nRMSE in counts are secondary measures of calibration performance. Over the three sets of segments used for calibration, the proposed approach improves the in-sample nRMSE in speeds by $41.6\%$, $68.4\%$, and $84.4\%$ compared to SPSA, and the in-sample nRMSE in counts by $46.1\%$, $63.8\%$, and $72.2\%$, respectively.

The performance of the proposed approach improves with the number of in-sample segments. The in-sample nRMSE in speeds decreases by $51.7\%$, $72.7\%$, and $89.2\%$, respectively, and the in-sample nRMSE in counts decreases by $51.0\%$, $71.5\%$, and $76.1\%$, respectively, compared to the speeds simulated using the initial demand. The out-of-sample nRMSE in speeds decreases by $51.6\%$ and $89.0\%$ over the 331 and 456 segment cases, and the out-of-sample nRMSE in counts decreases by $48.6\%$ and $71.8\%$, respectively. The percentage improvement over initial statistics increases with the number of segments used for calibration.
Out-of-sample nRMSE in speeds is smaller than in-sample nRMSE, which makes sense since out-of-sample segments mostly contain free-flow speeds that are easier to calibrate. The nRMSE in counts for the 331-segment case is large compared to acceptable levels observed in past work \cite{zhang2017efficient}, but it reduces to closer to acceptable levels with more segments, especially in-sample. Note that the nRMSE values in Table \ref{t:nrmse_insample_speeds} cannot be directly compared between segment sets because they contain different in-sample segments.

\section{Conclusion}
\label{s:conclusion}

This paper proposes an efficient metamodel-based approach for calibrating OD travel demands for urban road networks using GT speed data. We test the method under different data sparsity levels. Our numerical study shows that speeds-based calibration improves the fit to GT speeds by $89.2\%$ and the fit to GT counts by $76.1\%$ compared to the initial demands, with abundant data. It also shows the value of speed data, with increasing the number of segments with GT speed data improving the fit to both speeds and counts by $42.0\%$ and $48.8\%$, respectively. The proposed approach outperforms SPSA, a popular calibration algorithm, by $84.4\%$ and $72.2\%$ in terms of the fit to GT speeds and counts, respectively, under the same computational budget. Future work will investigate how abundant speed data can help mitigate the underdetermination (ill-posedness) of calibration problems, which is critical for using traffic models to perform counterfactually robust transportation analysis.

\bibliographystyle{ACM-Reference-Format}
\bibliography{sample-sigconf}

\end{document}